\documentclass[11pt]{article}
\usepackage{moriond,epsfig,amssymb}
\usepackage{hyperref} 

\def\be{\begin{equation}}
\def\ee{\end{equation}}
\def\bea{\begin{eqnarray}}
\def\eea{\end{eqnarray}}

\newcommand{\msq}{{m_{\widetilde{q}}}}
\newcommand{\mgo}{m_{\widetilde{g}}}
\newcommand{\mstau}{m_{\stau_1}}
\newcommand{\stau}{{\widetilde{\tau}}}
\newcommand{\mlcp}{m_\textnormal{\tiny LCP}}

\newcommand{\neu}{{\widetilde{\chi}^0_1}}
\newcommand{\mne}{{m_{\widetilde{\chi}^0}}}

\begin{document}
\vspace*{3.5cm}
\title{LONG-LIVED STAUS AT THE LHC}

\author{ JAN HEISIG }

\address{II.~Institute for Theoretical Physics, University of Hamburg\\
Germany
}

\maketitle\abstracts{
Supersymmetric scenarios with a very weakly interacting lightest 
superparticle (LSP)---like the gravitino or axino---naturally give 
rise to a long-lived next-to-LSP (NLSP). In the case of a stau NLSP,
the scenario shows up in a very prominent way at colliders. We present 
the LHC sensitivity for these classes of scenarios in the framework of
simplified models. This allows for deriving robust limits on the masses of
the lightest stau, the gluino and squarks. 
}

\section{Introduction}

In an $R$-parity conserving supersymmetric theory, the 
lightest supersymmetric particle (LSP) is stable and thus
provides us with a candidate for dark matter (DM).
Searching for supersymmetry at colliders always 
means searching for a certain LSP scenario. The neutralino 
LSP scenario is clearly the most widely studied scenario for 
the LHC and it is often referred to as \emph{the} SUSY search.
However, if one extends the MSSM, other cosmologically 
viable DM candidates show up. 
If, for instance,  we impose the idea of SUSY on
the fundamental theory of nature we have to supersymmetrize
gravity predicting the gravitino, the superpartner of the graviton. 
The gravitino is indeed a very well motivated LSP\,\cite{Fayet:1981sq} 
and may even be considered as favored since it alleviates the gravitino 
problem\,\cite{Ellis:1984er} allowing for larger reheating temperatures. 
Another perfectly viable DM candidate is the axino,\cite{Nilles:1981py}
a supersymmetric partner of the QCD axion 
which is introduced in the Peccei-Quinn mechanism.
Both candidates are very weakly interacting particles---the couplings 
are suppressed by the Planck- or the Peccei-Quinn scale, respectively. 
This naturally renders the next-to-LSP (NLSP) long-lived. 
If the decay length of the NLSP is large 
compared to the size of a detector of a collider experiment, the LSP is 
typically not involved in the interactions inside the detector at all and the 
NLSP determines the signature. For an electrically charged NLSP, like the stau, 
this leads to a very distinct SUSY signature---no missing energy 
but highly ionizing tracks leaving the detector.\cite{Fairbairn:2006gg}
Several studies for long-lived staus have been performed, assuming 
a certain underlying high scale model---such as minimal gauge mediated
supersymmetry breaking or the constrained MSSM (CMSSM).\footnote{%
Currently, searches for heavy stable charged particles are performed by
ATLAS\,\cite{Aad:2011hz} and CMS\,\cite{Chatrchyan1205.0272}.}
However, long-lived staus can basically occur in all common 
breaking models. 

In this work we address the question whether it is possible to derive robust bounds
on the mass parameters from the LHC experiment that cover all possible
spectra accommodating a long-lived stau NLSP. We will therefore adopt the 
approach of simplified models.\cite{Alwall:2008ag,Alves:2011wf} 
As an important outcome, we will see that in the case of long-lived staus
the simplified models provide a very powerful description covering a large set of
possible realistic spectra. We will restrict ourselves to the production via squarks 
and gluinos as well as to direct pair production of staus.

\section{Simplified models for long-lived staus}\label{sec:models}

Simplified models have successfully been used in the neutralino LSP 
scenario and have lead to bounds on the gluino and squark mass beyond 
those derived within the CMSSM.
However, one immediately notices some important differences between that
scenario and the one considered here. In the neutralino LSP scenario, the only
signal from the LSPs is missing energy. The SUSY search is almost 
independent of the neutralino kinematics, rather it depends strongly on the
type and hardness of the SM particle radiation. In the considered long-lived
stau scenario it is---roughly speaking---the other way around.
The kinematics of the staus, in particular their velocities are crucial quantities.  
For velocities well below one the signature is basically background free,
whereas, if the velocity approaches one a comparatively huge amount of
muon background is present. Consequently, the significance of the signal 
will drop drastically. 


When the stau is produced in a cascade decay following the 
production of colored sparticles a very large number
of parameters enter the computation through the
mixing angles and masses of the intermediate sparticles that 
can participate in the cascade decay. But not all parameters
are equally important. Obviously, the masses of the squarks
and gluinos play an important role in determining the production 
cross section. Additionally, the mass of the stau determines
the total phase space available in the cascade decay
of the lightest colored sparticle (LCP) down to the stau and 
thus dominantly determines the stau velocity. The impact
of the intermediate sparticles can successfully be described
by introducing a few distinct simplified models which represent
limiting cases of realistic spectra and allow for the free parameters
$\mgo$, $\msq$ and $\mstau$. 

Considering a fixed mass gap $\mlcp-\mstau$, 
in the limit of massless SM radiation a non-trivial extremum of 
the stau velocity is given by the mass pattern
\be
\label{eq:masspatt}
m_i=(\mlcp)^{\frac{n-i}{n}}(\mstau)^{\frac{i}{n}}
\ee
of $n-1$ intermediate sparticles $i$. This extremum is usually
a minimum. The only other extremum
is represented by the mass degenerate configurations
\be
\label{eq:massdeg}
m_i=\mlcp\,,\quad m_j=\mstau \quad 
\forall\; i<k,\; j\geq k\,, \quad k = 1,\dots,n \,.
\ee
which is kinematically equivalent to a direct decay of the LCP into 
the stau.
These observed extrema motivate the following simplified
models.
Model $\mathcal{A}$: The direct decay 
over a nearly degenerate neutralino ($\mne\simeq\mstau$).
Model $\mathcal{B}$:
The mass pattern (\ref{eq:masspatt}) for $n=2$,
LCP$\to\neu\to\stau_1$, i.e.
$\mne=\sqrt{\mlcp\mstau}$.
Model $\mathcal{C}$: The mass pattern (\ref{eq:masspatt}) 
for $n=4$, LCP$\to\widetilde{\chi}^0_2\to\widetilde{\ell}\to\neu\to\stau_1$.
In fact, these models represent limiting cases even for 
decay chains that radiate massive SM particles, like
 LCP$\to\widetilde{\chi}^0_2\to\neu\to\stau_1$. Decay chains longer than
 $n=4$ are rather unlikely and will not make up the dominant
 decay mode within the MSSM.\,\cite{Heisig:2012zq}


Another important difference between the neutralino LSP
and the considered scenario is that the latter provides a
perfectly well observable direct production mechanism.
The production via the Drell-Yan (DY) process depends only
on $\mstau$ and (less strongly) on the stau mixing angle $\theta_{\stau}$.
The stau mixing angle that provides the lowest cross section, 
$\theta_{\stau}^{\min}$,
is practically independent of the mass and kinematics of 
the stau (at least for stau masses above roughly $150\,\textnormal{GeV}$). 
Thus the DY production provides an important lower limit 
on the stau mass, which is completely 
model-independent.\,\cite{Heisig:2011dr} 

\begin{figure}[tbhp]
\begin{center}
\setlength{\unitlength}{1\textwidth}
\begin{picture}(0.85,0.74)
\put(0.0,0.4){
  \put(0.04,0.035){\includegraphics[scale=1.2]{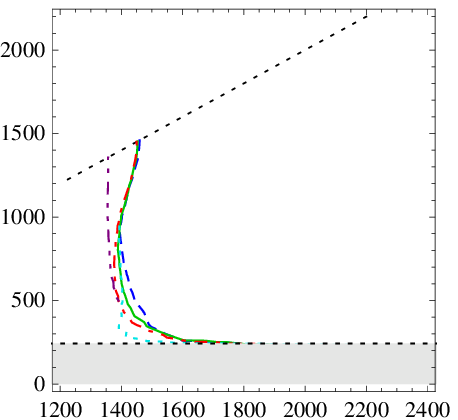}}
  \put(0.14,0.0){\footnotesize $\mgo=1/2 \,\msq\; [ \,\textnormal{GeV} \, ] $}
  \put(0.0,0.14){\rotatebox{90}{\footnotesize $\mstau\;[\,\textnormal{GeV}\,]$}}
  \put(0.09,0.316){\footnotesize $\,pp\,(8\,\textnormal{TeV})$}
  \put(0.09,0.286){\footnotesize $\int\!\mathcal{L}=16\,\textnormal{fb}^{-1}$}
  \put(0.136,0.073){\tiny minimal DY production limit}
  }
 \put(0.43,0.4){
  \put(0.04,0.035){\includegraphics[scale=1.2]{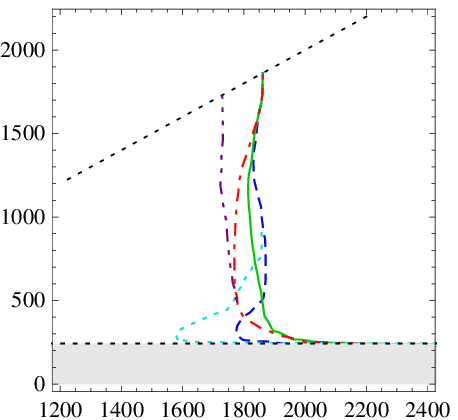}}   
  \put(0.15,0.0){\footnotesize $\mgo=\msq\; [ \,\textnormal{GeV} \, ] $}
  \put(0.0,0.14){\rotatebox{90}{\footnotesize $\mstau\;[\,\textnormal{GeV}\,]$}}
  \put(0.09,0.316){\footnotesize $\,pp\,(8\,\textnormal{TeV})$}
  \put(0.09,0.286){\footnotesize $\int\!\mathcal{L}=16\,\textnormal{fb}^{-1}$}
  \put(0.136,0.073){\tiny minimal DY production limit}
  }
\put(0.0,0.0){
  \put(0.04,0.035){\includegraphics[scale=1.2]{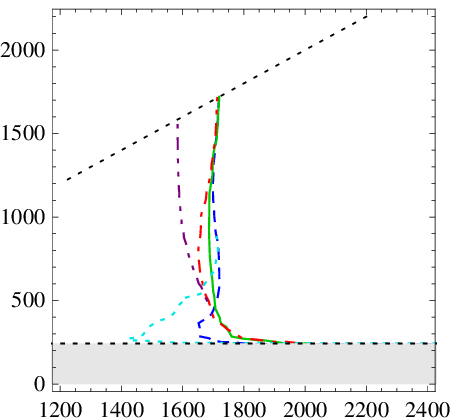}}
  \put(0.14,0.0){\footnotesize $\msq=1/2\, \mgo \; [ \,\textnormal{GeV} \, ] $}
  \put(0.0,0.14){\rotatebox{90}{\footnotesize $\mstau\;[\,\textnormal{GeV}\,]$}}
  \put(0.09,0.316){\footnotesize $\,pp\,(8\,\textnormal{TeV})$}
  \put(0.09,0.286){\footnotesize $\int\!\mathcal{L}=16\,\textnormal{fb}^{-1}$}
  \put(0.136,0.073){\tiny minimal DY production limit}
  }
 \put(0.43,0){
  \put(0.04,0.035){\includegraphics[scale=1.2]{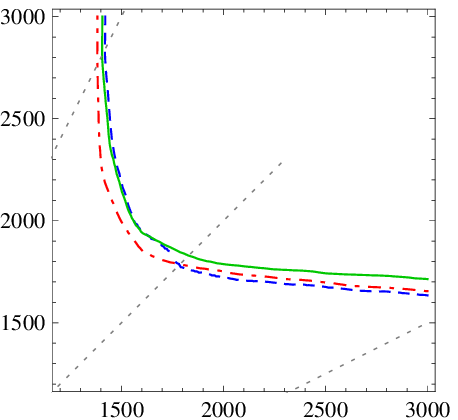}}   
  \put(0.18,0.0){\footnotesize $\mgo\; [ \,\textnormal{GeV} \, ] $}
  \put(0.0,0.14){\rotatebox{90}{\footnotesize $\msq\;[\,\textnormal{GeV}\,]$}}
  \put(0.215,0.316){\footnotesize $\,pp\,(8\,\textnormal{TeV})$}
  \put(0.215,0.286){\footnotesize $\int\!\mathcal{L}=16\,\textnormal{fb}^{-1}$}
 \put(0.215,0.26){\tiny along minima in}
 \put(0.215,0.244){\tiny $\mstau$-variation}
  }
\end{picture}
\end{center}
\caption{Projected LHC sensitivity ($95\%\,\textnormal{CL}_\textnormal{s}$ 
exclusion, see footnote \ref{footn}) for the models $\mathcal{A}$ (blue dashed), 
$\mathcal{B}$ (green solid) and $\mathcal{C}$ (red dot-dashed) as well as 
$\mathcal{A}$ (cyan dotted) and $\mathcal{C}$ (purple dot-dot-dashed) for
a reduced set of selection criteria (see text for details). In the lower right panel 
the curves represent the minima in the sensitivity with respect to the variation
of $\mstau$. We assume a common squark mass for all flavors. Taken from
reference 9.\,\,
}
\label{fig:8TeV}
\end{figure}

\begin{figure}[tbhp]
\begin{center}
\setlength{\unitlength}{1\textwidth}
\begin{picture}(0.85,0.36)
\put(0.0,0.0){
  \put(0.04,0.035){\includegraphics[scale=1.2]{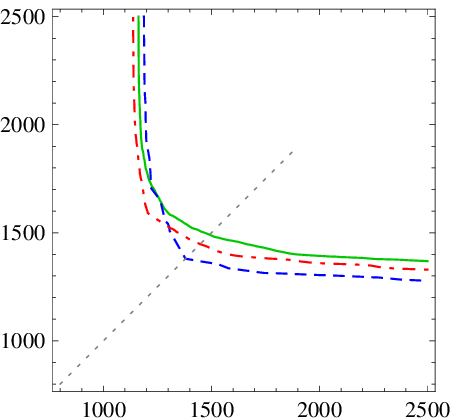}}   
  \put(0.18,0.0){\footnotesize $\mgo\; [ \,\textnormal{GeV} \, ] $}
  \put(0.0,0.14){\rotatebox{90}{\footnotesize $\msq\;[\,\textnormal{GeV}\,]$}}
  \put(0.21,0.316){\footnotesize $\,pp\,(7\,\textnormal{TeV})$}
  \put(0.21,0.286){\footnotesize $\int\!\mathcal{L}=5\,\textnormal{fb}^{-1}$}
 \put(0.215,0.26){\tiny along minima in}
 \put(0.215,0.244){\tiny $\mstau$-variation}
  }
 \put(0.43,0){
  \put(0.04,0.035){\includegraphics[scale=1.2]{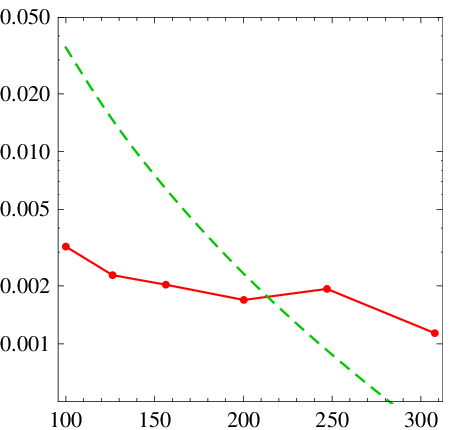}}   
  \put(0.18,0.0){\footnotesize $\mstau\;[\,\textnormal{GeV}\,]$}
  \put(0.0,0.17){\rotatebox{90}{\footnotesize $\sigma\; [ \,\textnormal{pb} \, ] $}}
  \put(0.215,0.316){\footnotesize $\,pp\,(7\,\textnormal{TeV})$}
  \put(0.215,0.286){\footnotesize $\int\!\mathcal{L}=5\,\textnormal{fb}^{-1}$}
 \put(0.215,0.256){\footnotesize minimal direct}
 \put(0.215,0.231){\footnotesize DY production}
 \put(0.1,0.324){\tiny theoretical}
  \put(0.11,0.308){\tiny prediction}
 \put(0.1,0.125){\tiny CMS limit}
 \put(0.1,0.108){\tiny Tracker+ToF}
  }
\end{picture}
\end{center}
\caption{Left panel: Projected LHC sensitivity 
($95\%\,\textnormal{CL}_\textnormal{s}$ exclusion, 
see footnote \ref{footn}) for the models $\mathcal{A}$ (blue dashed), 
$\mathcal{B}$ (green solid) and $\mathcal{C}$ (red dot-dashed). 
The curves represent the minima in the sensitivity w.r.t. the variation
of $\mstau$. We assume a common
squark mass for all flavors. Right panel: 
The cross section limit on stau pair production from the 
$7\,\textnormal{TeV}$ CMS data and the NLO
cross section prediction for the DY production with 
$\theta_{\stau}^{\min}$.
}
\label{fig:7TeV}
\end{figure}

\section{LHC reach}

To estimate the projected LHC reach of the scenario we
performed a full-fledged Monte Carlo study. We imposed a set of 
selection criteria which was optimized in order to provide
high efficiencies throughout the considered parameter space.
This was achieved by only taking into account the staus and 
up to two jets. The most important background is the DY production of
muons, even when two hard jets ($p_T>200\,\textnormal{GeV}$) are required.

Figure \ref{fig:8TeV} shows the projected $95\%\,\textnormal{CL}_\textnormal{s}$
exclusion limits\,\footnote{Due to the special situation of a very clean
signal region and thus only few required events these curves also represent an
(even conservative) estimation on the $5\sigma$ discovery potential.\label{footn}}
at the $8\,\textnormal{TeV}$ LHC for the three simplified models 
defined in section \ref{sec:models}.
In model $\mathcal{A}$ (blue dashed lines) the decay LCP$\to\neu$
is the dominant decay leading to hard jets and potentially highly boosted staus.
In an intermediate region of $\mlcp-\mstau$ this leads to an 
enhancement of the significance whereas for large mass gaps
$\mlcp-\mstau$ the significance drops sharply, despite the fact that the 
jets become harder. (The cyan dotted curve shows the sensitivity
without taking a jet-optimized selection criterium into account.)
The scenario would hide very effectively from observation if it
were not for the direct production which increases the sensitivity for
lower $\mstau$. This results in a minimum in the sensitivity
just above the limit provided by the minimal DY production.
For small mass gaps $\mlcp-\mstau$ the lower limit on the 
velocity becomes more important. Such a lower limit might be
imposed by trigger restrictions. The purple dot-dot-dashed
curves show the sensitivity for $\mathcal{C}$ after dropping
a selection criterium that allows staus to fire the muon trigger in delay,
i.e. requiring a buffering of the tracker data up to three bunch crossings
in delay.

In the lower right panel of figure \ref{fig:8TeV} and the left panel of figure
\ref{fig:7TeV} the LHC sensitivity in the $\mgo$-$\msq$-plane is shown. 
The curves show the $95\%\,\textnormal{CL}_\textnormal{s}$ taking the 
respective $\mstau$ that yields the lowest cross section at each point.
The simplified models $\mathcal{A}$, $\mathcal{B}$ and $\mathcal{C}$ 
span a relatively narrow band in the $\mgo$-$\msq$-plane. 
From the non-observation of heavy stable charged particles
this allows for an estimation of model-independent limits on 
$\mgo$ and $\msq$.
In the right panel of figure \ref{fig:7TeV} the cross section for direct DY 
production of staus is displayed as a function of the stau mass. We 
chose here the stau mixing angle which yields the smallest cross section. 
Additionally, we display the cross section limit curve from 
CMS\,\cite{Chatrchyan1205.0272} obtained for another choice of 
$\theta_{\stau}$. Since the kinematics do not depend on $\theta_{\stau}$, 
we do not expect any changes in the cut efficiencies.

\section{Conclusion}
The long-lived stau scenario can be very well described by simplified
models. The direct detectable stau does not give rise to regions
in parameter space where the scenario hides very effectively from 
observation, as it is e.g. the case in the neutralino LSP scenario where
compressed or widely spread spectra are difficult to explore. 
From the minimal direct production, long-lived staus with a mass below 
$\mstau=216\,\textnormal{GeV}$ can be excluded. This limit is completely 
model-independent. In the described framework, conservative limits 
on the gluino mass and a common squark mass of about $\mgo>1100\,\textnormal{GeV}$ 
and $\msq>1250\,\textnormal{GeV}$ can be estimated.

\section*{References}
\providecommand{\href}[2]{#2}\begingroup\raggedright\endgroup

\end{document}